\documentclass[12pt]{article}
\usepackage{graphicx,color}
\begin{document}

\begin{center}

{\large LARS BRINK AND SIC-POVMS}\footnote{For the memorial volume in honour 
of Lars Brink.}

\vspace{10mm}

{\large Ingemar Bengtsson}

\vspace{7mm}

{\sl Stockholms Universitet, AlbaNova\\
Fysikum\\
S-106 91 Stockholm, Sverige}

\vspace{5mm}

{\bf Abstract}: 

\end{center}

{\small

\noindent The notion of SIC-POVMs comes from quantum theory, and they were not on the 
horizon when I was Lars Brink's student in the early 80s. In the summer of 2022 I told Lars 
that I know how to use number theoretical insights to construct SIC-POVMs in any Hilbert 
space of dimension $n^2+3$, and that the construction provides a geometric setting for 
some deep number theoretical conjectures. I will give a sketch of this development, of 
what it was like to be Lars' student, and of what his reaction to our construction was. 
}

\vspace{8mm}

{\bf 1. Introduction}

\

\noindent This contribution splits into two somewhat disparate parts: An account of 
what it was like to be one of Lars Brink's many students, and a review of some 
really pleasing developments in the SIC-POVM existence problem. Lars needs no 
introduction to the readers of this volume, but SIC-POVMs probably do. First of all 
I prefer to call them SICs, and to treat ``SIC'' as a noun rather than as an 
acronym. To define a SIC, consider a 
Hilbert space of complex dimension $d$. We can view the pure quantum states as density 
matrices forming a small submanifold of the surface of a ball in a real space of 
dimension $d^2- 1$. This is just the embedding of complex projective space into the 
set of Hermitean matrices of fixed trace \cite{BZ}. A SIC is a maximal regular simplex in 
the latter, placed in such a way that all its $d^2$ vertices are pure quantum states. 
The question---which does have a practical quantum engineering aspect that I will not 
go into---is whether this can be arranged in any dimension, or not? The conjecture 
is that it can, 
and that these vertices form an orbit under a Weyl--Heisenberg group \cite{Zauner, Renes}. 
However, proving the conjecture, and actually constructing SICs when the dimension is 
higher than two, has proven very difficult. The $d = 3$ case was done by Otto Hesse 
in 1844, but until recently constructions were limited to dimensions that are only a 
factor of ten or so higher than that. We now have a well-defined recipe for how to do 
it that we believe will work for every dimension of the form $d = n^2+3$, and which we 
have actually carried through in something like seventy different cases, the highest 
being $d = 39604$. 

The way it works can be guessed if we consider a simpler problem, that of constructing 
regular polygons in the plane. Their vertices are easily described using roots of 
unity, that is to say by extending the rational number field to the cyclotomic 
number fields. In this way the regular polygons are to be found at the intersection 
of geometry with an important---if by now well known---chapter of number theory. 
Constructing SICs works in a somewhat similar way. But they sit at an 
unexpected intersection of geometry with a chapter of number theory that is partly 
unwritten at present. This is what makes it exciting. I will return to it 
in sections 3 and 4, but first I will devote section 2 to some memories about what it was 
like to be Lars' student. In section 5 I join the two different strands 
together. 

\

\

{\bf 2. The group in G\"oteborg}

\

\noindent Having been admitted as a graduate student by the Institute of Theoretical 
Physics in G\"oteborg in the autumn of 1979, I decided to join the elementary particle 
physics group after hearing Lars Brink give a sketch of what he was working on at the 
time. Although I did not know the background, this was when particle physics was still 
flushed with the successes of the early 70s. The idea was to unify all forces and all matter 
in what was, on group theoretical grounds, the largest framework possible. It was called 
$N = 8$ supergravity \cite{Brink1}. This theory should, incidentally, also solve the problem of 
how to quantize gravity. Since I had heard that this was an unsolved problem, this sounded 
interesting. Of course, despite his friendliness, Lars was also a little intimidating: 
somehow I got hold of his PhD thesis \cite{Brink2}, and realized that it would be hard to 
write an equally good one. 

I was given a desk, and was told where one could get pen and paper. Since smoking was not 
allowed in the offices it took a little time before I came to the department regularly, 
but I soon discovered what a wonderful place it was. The group consisted of Lars Brink, 
Robert Marnelius, Arne Kihlberg, Ulf Ottoson who was the best physics teacher I have 
ever come across, Giorgio Peressutti who managed all the incoming preprints, a crowd of 
students, Jan Nilsson who was a somewhat distant but benign Professor, and Britta Winnberg 
who was a superbly competent secretary and typed all our papers. We had exciting 
discussions at lunch, either at the University canteen or (when Lars got his way) at a 
Chinese restaurant called Tai Pak. Outside lunch hour Lars' office was always open, should 
one have a question to ask, and every Friday there was a two hours long seminar where 
a member of the group explained something of interest. If the speaker was a student this 
was usually some paper that he had read.

\begin{figure}[t]
\center{
\includegraphics[angle=0,width=0.50\columnwidth]{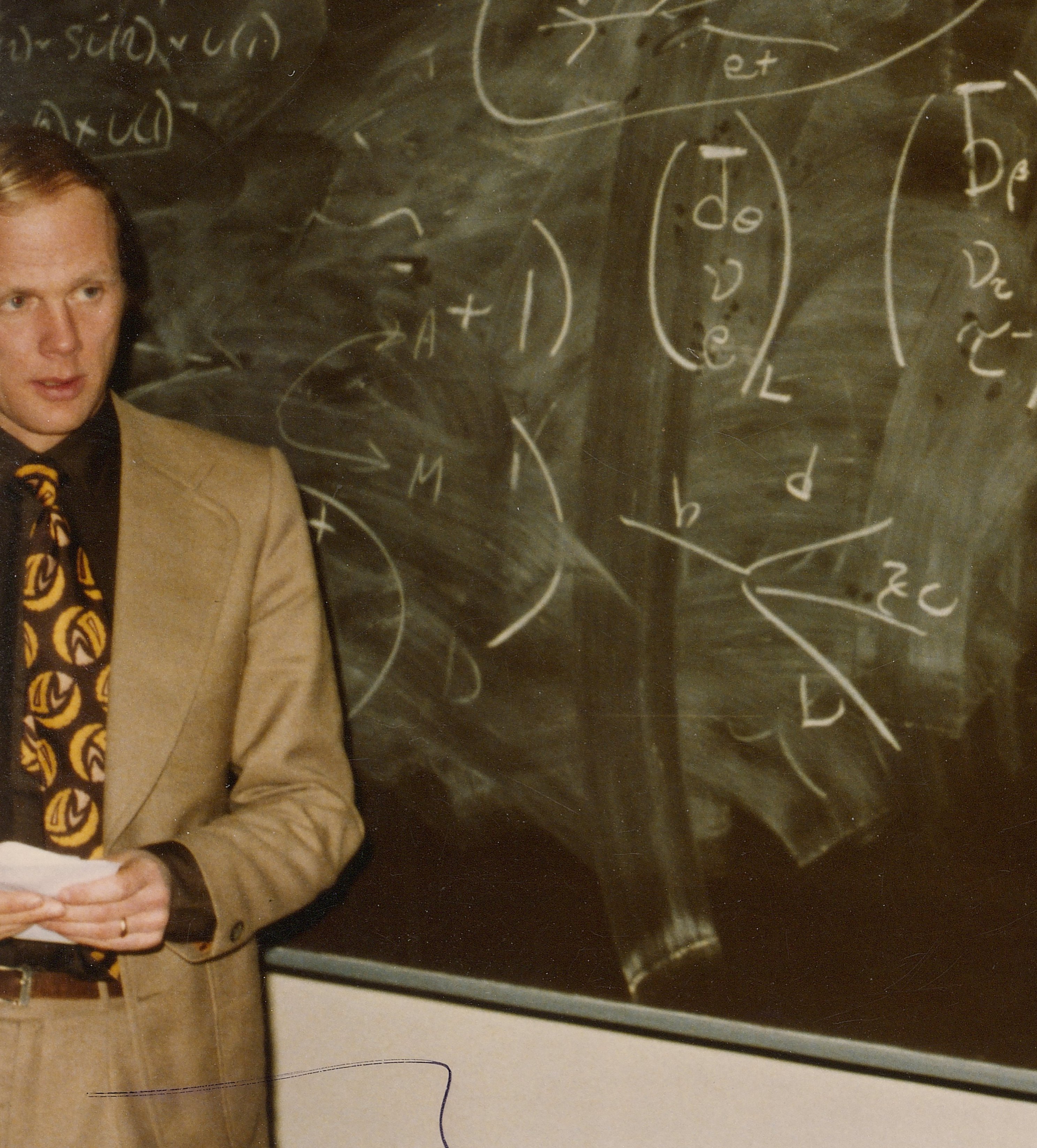}}
\caption{Lars Brink in 1979. The scribblings on the board are due to Sheldon Glashow, hence the tie.}
	\label{fig:3}
\end{figure}

At the beginning of each term Lars used to hold a soliloquy about what we ought to do. 
The trick was to find problems that were relevant, and to which we could actually 
contribute. When things did not work Bengt Nilsson, who was the older student, 
used to comfort me by saying that eventually all would be well if I only kept on calculating. 
Keeping these three lessons in mind has been very helpful over the years. 

At first I was introduced to the problem of formalizing the notion of a point particle, 
building on the papers that Lars had written a few years earlier \cite{Brink3, Brink4}, 
and utilizing Dirac's theory of constrained Hamiltonian systems heavily. But Lars used 
to go to Caltech once a year to work with John Schwarz, and he always came back with 
the most fantastic ideas. From 1981 and on we knew that $N = 8$ supergravity was not the 
answer to everything. The theory to end all other theories was known as Superstring Theory 
\cite{Brink5}. Moreover, this theory 
was to be formulated in the lightcone gauge, using ideas that again went back to Dirac. 
It was great to attend conferences and summer schools and know that what we were doing 
in our own group was actually much more to the point than what was presented in the 
lectures. 

Of course our interests were not confined to superstring theory. One of Lars' students worked 
on Bell's theorem, a subject that was totally obscure at the time, and John Bell came to 
visit in order to help with this work \cite{Ogren}. So we had many discussions about the 
foundations of quantum mechanics. Moreover, following Polyakov's brilliant use of the 
Brink--Di Vecchia--Howe string action, Robert and Arne wrote fine papers on the 
Liouville theory. I did not understand it at the time, but it will resurface below. 

By 1983 we had grasped the idea that the light cone gauge formulation can stand on its 
own legs, without any covariant crutches \cite{Brink6}. Following discussions at Tai Pak we tried to 
construct a field theory for higher spins in this way, and wrote what I think were the most 
exciting papers that I contributed to myself \cite{BBB1, BBB2}. In 1984 I wrote my thesis, which 
as I had predicted was not as good as Lars', but was good enough to earn me a CERN fellowship. 

Just before I came to Geneva, 
Mike Green and John Schwarz had discovered the anomaly cancellation mechanism in chiral 
superstrings, and the world turned upside down. Suddenly everybody agreed that superstrings 
held the answers to all questions. I rather preferred the way the world had been when 
it was only us who thought so, and gradually drifted into other subjects. Subjects that have 
now been flushed with success, such as classical general relativity, and quantum information 
theory. The problem to which we turn in the next section comes from quantum information 
theory, but has not really stayed there. 

\

\

{\bf 3. The SIC existence problem and the Stark conjectures}

\

\noindent The SIC existence problem is easy to state. We want to find sets of $N$ 
unit vectors in the finite dimensional Hilbert space ${\bf C}^d$, and constants 
$c_1$ and $c_2$, such that 

\begin{equation} \sum_{i=1}^N|\psi_i\rangle \langle \psi_i| = c_1 {\bf 1} \label{tightframe} 
\end{equation}

\begin{equation} |\langle \psi_i|\psi_j\rangle |^2 = c_2 \hspace{5mm} \mbox{if} 
\hspace{5mm} i \neq j \ . \end{equation}

\noindent In the signal processing community such sets are known as equiangular 
tight frames. They can 
be thought of as $N$ points evenly spread in complex projective space, or as a regular 
simplex in the space containing the convex set of all mixed quantum states, carefully 
centred and arranged so that all its corners are pure. When thinking projectively 
one may refer to the vectors as representatives of $N$ equiangular lines in Hilbert 
space. One proves easily that if the arrangement can be done at all then 

\begin{equation} d \leq N \leq d^2 \ , \hspace{10mm} 
c_1 = \frac{N}{d} \ , \hspace{10mm} c_2 = \frac{N-d}{d(N-1)} \ . \end{equation}

\noindent If $N = d$ we have an orthonormal basis, and they clearly exist for all $d$. 
If $N = d^2$ we have a SIC, also known as a maximal equiangular tight frame. 
This time existence is not a foregone conclusion. In 
fact, in dimension three the possible values of $N$ are 3, 4, 6, 7, and 9, while $N = 5$ and 
$N = 8$ are impossible \cite{Feri}. And this is about as far as one gets with unaided 
Gr\"obner basis techniques. 

In the SIC existence problem we ask if SICs exist in any dimension $d$. The way I 
have presented it here is the way I first presented it to Lars, and it makes the problem 
sound like an interesting hobby problem at most. Indeed this was Lars' first reaction. 
He eventually changed his mind about this, but to see why we have to see where the 
problem leads. 

Some ideas are needed. The first idea is that SICs---like orthonormal bases---are 
orbits of a group, and more precisely that they are orbits of a Weyl--Heisenberg 
group having an essentially unique unitary representation in ${\bf C}^d$. This means 
that, up to phase factors, we have $d^2$ group elements $D_{i,j}$ where $i,j$ are 
integers modulo $d$, and the $d^2$ vectors in the SIC take the form 

\begin{equation} \{ |\Psi_{i,j}\rangle \}_{i,j = 0}^{d-1} = \{ 
D_{i,j}|\Psi_0\rangle \}_{i,j = 0}^{d-1} \ , \end{equation}

\noindent where $|\Psi_0\rangle$ is known as the fiducial vector for the SIC. 
Numerical searches then become a realistic option, and indeed numerical SICs 
have been found for all $d \leq 193$, although the searches remain 
non-trivial. We are talking about CPU years here \cite{Andrew, Grassl}. An added 
advantage is that the group provides us with a preferred basis in Hilbert 
space \cite{Weyl}, which means that we can discuss number theoretical properties 
of the components of the fiducial vector, should we wish to do so. 

The next idea is to ask for symmetries of the SICs, that is to say unitary 
operators that leave the fiducial vector invariant.\footnote{I 
warn the reader that in this paragraph I oversimplify the story just 
a little bit. In particular, here and elsewhere I ignore the distinction between 
even and odd dimensions that runs through the subject.} Such operators should belong 
to the automorphism group of the Weyl--Heisenberg group, which turns out to mean 
that they represent the group $SL(2, {\bf Z}_d)$, that is to say to 
symplectic group over the integers modulo $d$, with the representation fixed 
once the representation of the Weyl--Heisenberg group has been fixed. The pair of 
indices on the group elements $D_{i,j}$ can be thought of as vectors on which 
the symplectic matrices act \cite{Marcus}. With 
very little evidence, Zauner \cite{Zauner} suggested that SICs do indeed 
exhibit such a symmetry, and that the order of the symmetry group is divisible 
by three. This has been confirmed for all SICs found in dimensions $d \leq 50$ \cite{Scott}, 
and has since then been built into the numerical searches in higher dimensions. Importantly, 
if the dimension $d$ is of the special form $d = n^2+3$ then 
anti-unitary symmetries also appear, which means 
that we deal with $GL$ matrices of determinant $\pm 1$ modulo $d$. 

We need more ideas. Explicit exact solutions for the SIC fiducial vectors do 
not look pleasant \cite{Scott}, but in a sense they are. One finds that they 
can always (unless $d = 3$) be expressed in terms of (very) nested radicals, 
which is surprising since the defining equations are multivariate polynomial 
equations that would not---as we learned from Abel and Galois---be expected 
to have any such solutions. A possible explanation for this was found ten 
years ago \cite{AYAZ}. The claim is that, for some reason, the numbers that 
enter the fiducial vector always belong to number fields that are extensions of a 
real quadratic number field ${\bf Q}(\sqrt{D})$, where $D$ is a positive integer and 
the Galois group of this extension is abelian. Moreover the quadratic field can 
be related to the dimension $d$ in which we are constructing 
the SIC, through the magical formula 

\begin{equation} D = \ \mbox{square-free part of} \ (d+1)(d-3) \ . 
\label{magi} \end{equation}

\noindent We do not yet understand that, but by now the evidence for the claim 
is overwhelming. 

But to get a SIC we need an extension of this number field, with abelian 
Galois group. And this is where we plunge into deep waters. Let us first ask, 
what are the abelian extensions of the rational number field ${\bf Q}$? The 
answer was given long ago by Kronecker and Weber, and says that any abelian 
extension of the rational field is a subfield of some cyclotomic number field, 
that is a number field obtained from the rational field by adjoining an 
$n$th root of unity for some choice of the integer $n$. Moreover any root 
of unity can be obtained by evaluating the analytic function 

\begin{equation} e(x) = e^{2\pi ix} \end{equation} 

\noindent at a rational point. There is a primitive $n$th root of unity $\omega_n$ 
of the form  

\begin{equation} \omega_n = e(1/n) \ . \end{equation}

\noindent And every abelian extension of ${\bf Q}$ is a subfield of 
${\bf Q}(\omega_n)$ for some choice of $n$. Interestingly, the cyclotomic 
fields are all you need in order to write down the unitary representations 
of the Weyl--Heisenberg groups and their automorphism groups. Just choose 
$n = d$, where $d$ is the Hilbert space dimension you are interested in. 

In his famous 1900 address to the mathematical community, Hilbert asked 
for a similar description of the abelian extensions of arbitrary number fields 
\cite{Hilbert}. However, after more than a hundred years of work, number 
theorists have only partial results if the base field that you extend is 
the real quadratic field ${\bf Q}(\sqrt{D})$. The extensions have been 
classified, and depend on choosing an ideal among the integers in the 
base field. Given such an ideal one can construct an abelian ray class group, 
and then there is a theorem guaranteeing the existence of an extension 
of the base field whose Galois group is isomorphic to the ray class group. 
This is known as the ray class field with 
the given ideal as its modulus. There are also algorithms implemented in 
computer algebra packages that allows one to construct these ray class fields, 
provided that their degrees are not too large. But an elegant description, 
using some analogue of the analytic function $e(x)$, has not yet been 
found. 

The available information is however enough to formulate a sharp conjecture about 
SICs. Known as the AFMY conjecture \cite{AFMY}, it states that in every dimension 
$d$ there exists a SIC that can be constructed using the ray class field with the 
above real quadratic field as its base field, and modulus equal to $d$. 
Moreover this is the smallest number field that can be used for the purpose. 

The analogue of the function $e(x)$ has so far eluded number theorists, and the 
AFMY conjecture by itself is not enough to construct SICs. Here the Stark conjectures, 
from the 1970s, come in \cite{Stark}. Just as Riemann's zeta function $\zeta (s)$ encodes 
arithmetic information about the ordinary integers, there are $L$-functions that 
encode information about other number fields. For each element $\sigma$ in the ray 
class group Stark formed an analytic function $\delta (s,\sigma )$ from partial 
$L$-functions, and conjectured that an algebraic unit can be found by evaluating its 
derivative at $s = 0$. In favourable cases the Stark units generate the ray class field. With a suitable 
embedding of the number field one can in principle calculate, to any desired precision, 
such a Stark unit $\epsilon_\sigma$ from the formula 

\begin{equation} \epsilon_\sigma = e^{\delta^\prime (0,\sigma)} \ . \end{equation}

\noindent Since I have not given the definition of the function $\delta$ this hangs 
in the air unless you consult the references, but I want to draw attention to the fact 
that the Stark units so defined have a sign, which makes it natural to ask if their 
square roots belong to the same number field. These Stark units are then analogous 
to the roots of unity for the ray class fields we are interested in, 
and they form Stark's conjectural answer to Hilbert's question. The analogy was 
strengthened by Shintani, who found a way to express the right hand side as a product 
of special values of Barnes' double gamma function \cite{Shintani}. 

Can we use Stark units to construct SICs? Gene Kopp did just that, in dimensions 
5, 11, 17, and 23 \cite{Kopp}. In the work I took part in we followed a rather different 
track, taking us to much higher dimensions. 

\

\

{\bf 4. Constructing SICs from Stark units}

\

\noindent Finally I come to our construction. I will be a bit sparing with references 
because the paper we wrote has many \cite{PaperI}. The authors---Marcus Appleby, 
Markus Grassl, Michael Harrison, Gary McConnell, and myself---include two number theorists, 
while three of us come from quantum information theory. The work is cross-disciplinary, 
because it has to be. 

We rely on a string of conjectures to get started, but at the end we check 
if our candidate SIC fiducial vector defines a SIC, or not. In the close 
to seventy cases that we have checked, it always is. 

Let the dimension be of the form $n^2+3$. Elementary number theory tells us that 
its prime decomposition must be \cite{units} 

\begin{equation} d = n^2+3 = 4^{e_0}\cdot 3^{e_1}\cdot p_1^{k_1} \cdot \dots 
\cdot p_r^{k_r} \ , \end{equation}

\noindent where $e_0,e_1 = 0$ or 1 and all the remaining primes are equal to 
1 modulo 3. The last fact is highly significant for the represention theory of 
the symmetry group of the SIC, as we will see. On the number theory side it turns 
out that the real quadratic fields that turn up for these dimensions, via the 
magical formula (\ref{magi}), are precisely those that admit a unit of negative 
norm. So we are dealing with a quite exceptional series of dimensions. 

I will focus on the case $d = p$ here. With some extra work all other cases can be 
covered using the same ideas but there are some complications. When the dimension 
is composite we have to deal with an entire lattice of subfields, and moreover 
a lovely trick \cite{monomial} has to be brought in if the dimension contains a 
factor of four. This would take some time to explain so, in the interests of brevity, 

\begin{equation} d = n^2+3 = p \hspace{5mm} \Rightarrow \hspace{5mm} p = 1 \ 
\mbox{modulo} \ 3 \ . \end{equation}

\noindent According to a long standing conjecture by Hardy and Littlewood this is 
an infinite sequence in itself. Now, why is it significant that the prime equals 
one modulo three? The answer has to do with the expected symmetry group of the SICs. 
We expect it to represent some subgroup of $SL(2,{\bf Z}_p)$ of order divisible 
by $3\ell$, where $\ell$ is an integer that can be calculated if the dimension is 
known \cite{AFMY}. What we have to know about the representation theory is that 
most $SL$ matrices are represented by complex Hadamard matrices all of whose matrix 
elements are roots of unity up to a factor, but diagonal $SL$ matrices are 
represented by real permutation matrices \cite{Marcus}. Now the multiplicative 
group ${\bf Z}_p^\times$ contains primitive elements 
$\theta$ of order $p-1$. Let us consider the diagonal matrix 

\begin{equation} C = \left( \begin{array}{cc} \theta^{-1} & 0 \\ 
0 & \theta \end{array} \right) \ , \hspace{8mm} \theta^{p-1} = 1 \ \mbox{mod} \ p 
\ . \end{equation} 

\noindent If $p-1$ is divisible by 3 the symmetries can be generated by 
$C^{(p-1)/3\ell}$, which has order $3\ell$. The great thing with this 
is that, in these dimensions, we can arrange the symmetry group so that it simply 
permutes the components of the vector. 

If, in addition, the dimension is equal to $n^2+3$ the standard conjectures say that 
there exists a SIC fiducial vector that is invariant under complex conjugation as 
well \cite{Andrew}. If we apply the discrete Fourier 
transform to such a SIC vector we arrive at a fiducial vector for another SIC, unitarily 
equivalent to the first one, and an argument originally due to Einstein \cite{Einstein} 
implies that this fiducial vector takes the almost flat form 

\begin{eqnarray} \Psi = (a_0, a_1, \dots , a_{d-1})^{\rm T} \ , \hspace{25mm} \nonumber \\ 
\label{fid} \\ 
\hspace{1mm} |a_1| = |a_2| = \dots = |a_{d-1}| \ , \hspace{5mm} |a_0|^2 = (2+\sqrt{d+1})|a_1|^2 
\ .  \nonumber \end{eqnarray}

\noindent It is almost flat in the sense that all components but one have the same 
absolute value. The permutation matrix representing the symplectic matrix $C$ leaves 
the first component invariant and permutes the rest among each other. Because the 
matrix $C^{(p-1)/3\ell}$ gives rise to a symmetry this means that there are only 
$(p-1)/3\ell$ independent numbers in the vector. 

The permutation matrix representing the matrix $C$ acts on the vector in a very nice way. 
At the moment its components are indexed by an integer $j$ modulo $p$. The integer $\theta$ 
that occurs in the matrix is a primitive element in the ring of integers modulo $p$, which 
means that for each non-zero integer $j$ we can find an integer $r$ such that 

\begin{equation} j = \theta^r \ . \label{primitive} \end{equation}

\noindent The permutation matrix then acts in such a way that $a_{\theta^r} \rightarrow 
a_{{\theta}^{r+1}}$. Which brings us to a third conjecture: The Galois group of the number 
field out of which the vector is constructed acts in the same way. So, in this way, 
the whole vector is generated from the single number $a_1$. Perhaps it is a Stark unit? 

Now recall the magical formula (\ref{magi}) for the real quadratic base field 
$K = {\bf Q}(\sqrt{D})$. Because $d-3 = n^2$ 
the square root $\sqrt{d+1}$ belongs to the base field. Moreover 

\begin{equation} d = (\sqrt{d+1}+1)(\sqrt{d-1} - 1) \ . \end{equation}

\noindent It is easily checked that the factors on the right hand side are algebraic integers. 
Hence the rational prime $d = p$ does not remain prime over $K$. We can use any one of the 
factors to define the modulus of a ray class field which will be a subfield of the full 
SIC field. Denoting the factors by $\partial$ and $\bar{\partial}$ we get two smaller ray 
class fields that we denote by $K^{\partial, [1]}$ and $K^{\bar{\partial}, [2]}$, while the 
large ray class field containing the SIC is $K^{d,[1,2]}$. They are all extensions of the 
real quadratic field $K$, and their degrees are found to be 

\begin{equation} {\rm deg}\left( K^{\partial,[1]}\right) = {\rm deg}\left( 
K^{\bar{\partial},[2]}\right) = \frac{h(p-1)}{3\ell} \ , \hspace{4mm} {\rm deg}\left( 
K^{d,[1,2]}\right) = \frac{h(p-1)^2}{3\ell} \ . \end{equation}

\noindent Here $h$ is the class number of $K$, that is to say the degree of Hilbert's 
class field over $K$.\footnote{And the little brackets have to do with ramification, and 
how these number fields can be embedded into the complex field. I do not go into this 
here, nor will I explain the role of Hilbert's class field $H$.} Hence, if the fiducial 
vector can be constructed using only the 
number field $K^{\partial,[1]}$, say, there will be $h(p-1)/3\ell$ Stark units, which is 
just enough to construct $h$ unitarily inequivalent almost flat SIC fiducial vectors. 

We still need the large ray class field $K^{d,[1,2]}$ to construct the SIC. What happens 
is that the smaller ray class fields have trivial intersection with the cyclotomic 
field ${\bf Q}(\omega_d)$. But we need the roots of unity to represent the Weyl--Heisenberg 
group, which means that the larger ray class field will be filled out once we have 
acted on the fiducial vector with the group in order to construct all the $d^2$ vectors 
in the SIC. 

With a suitable embedding of our number fields into the field of complex numbers one finds 
that $K^{\partial,[1]}$ contains real numbers only, while the numbers in $K^{\bar{\partial},[2]}$ 
are complex. The Stark units are calculated in the real branch of the field. They can be 
taken over to the complex field $K^{\bar{\partial},[2]}$ by means of the Galois transformation 
$\sqrt{D} \rightarrow - \sqrt{D}$, and are then found to sit on the unit circle in the 
complex plane. In other words, they are phase factors. 

Since we want to build a SIC fiducial vector from Stark units it seems natural to aim for 
a real fiducial vector once we have calculated the real Stark units. However, although 
units certainly play a role there \cite{units}, it did not work when we tried it. So we 
take the Stark units over to the complex branch. If 
we normalize the almost flat fiducial vector (\ref{fid}) so that the components $a_1, a_2, 
\dots , a_{d-1}$ are phase factors, we can ask if they are also Stark units? But this would 
disagree with the AFMY conjecture \cite{AFMY} because with this normalization the first component $a_0$ 
does not belong to the ray class field. We can renormalize the vector so that the first 
component has absolute value $2 + \sqrt{d+1}$. This is in the field, but then the remaining 
components are not phase factors anymore. Instead they will take the form 

\begin{equation} a_j = \sqrt{(-2-\sqrt{d+1})e^{i\vartheta_j}} \label{sqrt} \end{equation}

\noindent for some phase factors $e^{i\vartheta_j}$. Examination of a few low dimensional 
cases showed that, indeed, these phase factors are Stark units. Eventually we proved 
that, although the square roots of these Stark units do not belong to the ray class field, 
the rescaled square roots---as in equation (\ref{sqrt})---do. That is, we have 
a procedure for how to calculate SICs from first principles!

First we calculate the real Stark units from the $L$-functions. They come out as an ordered 
set of real numbers in decimal form. The ordering is provided by the ray class group, which 
is isomorphic to the Galois group of the field extension. We increase the precision until 
we can determine the minimal polynomial of the units, and then perform the Galois 
transformation to the complex branch of the field. With the minimal polynomial in hand 
we can rescale its roots and take their square roots with only one overall sign left 
ambiguous. With some thought, and some calculation, the ordering of the real roots can 
be taken over to the 
complex roots. We then have the correct number of ordered numbers to place in our 
fiducial vector, and it only remains to check that it is, in fact, a SIC fiducial vector. 

While the idea is simple, there are some hurdles to overcome. The main difficulty is 
that calculating the real Stark units to the precision needed is a time consuming task. 
The published version of our procedure \cite{PaperI} also has some blemishes. When we 
wrote it we did not know how to match the cyclic action of the Galois group to the 
cyclic ordering provided by the symmetry group of the SIC, and had to perform a search 
over all possible choices of the primitive element $\theta$ in equation (\ref{primitive}). 
This has now been resolved, and will be published sometime. By now we also have a 
simple way to choose the global sign that appears when we take the square roots, 
so we can claim that our procedure is fully deterministic in the prime case. And it works in every 
case we have tried, including all dimensions $d = n^2+3$ with $n\leq 53$ and 15 
higher dimensional cases. 

Looking back over the argument, we can draw a few conclusions. First, every conjecture 
that we relied on holds, in all the dimensions where we were 
able to carry through the calculations. Second, we have 
found a new role for the Stark units. At some level they could always be thought of 
as analogous to the roots of unity that generate the cyclotomic 
number fields. But the whole {\it raison d'\^{e}tre} of the roots of unity is that they 
divide the circle into equal parts, that is to say that they solve a geometrical 
problem. We have found that the Stark units also solve a geometrical problem, namely 
that of finding maximal sets of equiangular lines in complex Hilbert spaces. This seems 
to be a highly significant fact, even if it is slightly dimmed by the fact that we 
do not really understand why it happens. Lars told me that I should continue 
to work on this problem, and I will take his advice. 

\

\

{\bf 5. Envoi}

\

\noindent All good stories come to an end, and in late May 2022 another of Lars' 
students, Anna Tollst\'en, told me that Lars was very ill. I wrote emails to him, but although 
I understood that he was in pain I received only cheerful and interesting replies. He was 
reading up on the life of Einstein at the time, and when I told him that I had cited 
one of Einstein's technical papers on the relation between the power spectrum and the 
autocorrelation function \cite{Einstein} he was a little bit envious. He himself 
never cited Einstein. It is good for a student 
to feel that he has impressed his supervisor, at least somehow. 

When my collaborator Marcus Appleby visited me in August, for a week at the blackboard, 
I discovered that Marcus was at one stage a student of Mike Green's, so 
maybe there was something about what Lars and Mike were doing then that led naturally to SICs. 
In fact, what Marcus and I were discussing at the board was Faddeev's quantum 
dilogarithm, a most special function that appears in Liouville theory \cite{Faddeev}, and 
had just made its appearance in the SIC problem. So I told Lars that I regretted that 
I had not listened carefully enough when he and Robert told me what an interesting 
quantum field theory that is. Swedish readers will understand what Lars meant when he 
replied that I should ``give the iron'' with SICs. 

\

{\small

\end{document}